# Efficient Third-Harmonic Generation via Strong Coupling of Quasi-Guided Modes

Xin Huang,[1, 2] Qi Lin,[1, 3] Sheng-Xuan Xia,[3] Xiang Zhai,[3] Gui-Dong Liu [1, 3, *]

[1] *School of Physics and Optoelectronics, Xiangtan University, Xiangtan 411105, China*

[2] *Hunan Engineering Laboratory for Microelectronics, Optoelectronics and System on a Chip, Xiangtan University, Xiangtan 411105, China*

[3] *School of Physics and Electronics, Hunan University, Changsha 410082, China*

*[gdliu@xtu.edu.cn](gdliu@xtu.edu.cn)

**Abstract:** Improving the conversion efficiency is critical for advancing nonlinear photonic devices, enabling applications in ultrafast optics, quantum light sources, and high-resolution imaging. Herein, we present a bilayer waveguide integrated with a periodic square nanocube array to enhance third-harmonic generation (THG) efficiency. This structure achieves strong coupling between TE- and TM-polarized quasi-guided mode (QGM) resonances, thereby enabling simultaneous dynamic control of both resonant *Q*-factor and coupling strength through incident polarization angle. The distinct avoided crossing observed in both reflection spectra and eigenfrequency diagrams indicates the strong coupling regime. This mechanism, facilitated by the inter-mode energy exchange within the hybridized system, results in an enhancement of *Q*-factor by two orders of magnitude when compared to conventional guided-mode resonances. Simulation results verify an unprecedented maximum *Q*-factor of $10^{12}$ for the upper-branch mode. Leveraging strong-coupling-induced field enhancement, the system attains THG conversion efficiency of order $10^{-2}$, demonstrating unprecedented nonlinear performance.

## 1. Introduction

High-Q photonic nano-resonators serve as core components in all-dielectric nanophotonics and constitute essential elements for high-performance optoelectronic and photonic devices in modern optical communications. They find applications in filters [1], sensors [2], and nonlinear optics [3]. Strategies for achieving high-Q resonances primarily include two approaches: QGM resonance [4,5] and QBIC [6-8]. Compared to QBIC, QGM resonance exhibits significant advantages. First, enhanced near-field performance is achieved by QGM resonance, benefiting from the properties of its precursor, the GM. This results in stronger near-field localization enhancement and higher excitation efficiency, which is critical for applications requiring intensified light-matter interactions [9,10]. Second, QGM resonance demonstrates robust angular tolerance. The *Q* factor of QGM resonance remains exceptionally robust across a wide wave vector range, corresponding to large incident angle conditions [11]. This characteristic makes it particularly suitable for nonlinear optical processes that utilize incident angle for spectral tuning. The resonance frequency can be adjusted while the *Q*-factor and the corresponding field enhancement effect are effectively maintained.

However, single QGM resonance still exhibits certain limitations, which can be overcome by the hybrid modes formed through coupling. The strong coupling mechanism breaks the independence of orthogonal modes (TM/TE), establishes a dynamic energy exchange channel (Rabi oscillations) between them, and forms hybrid electric fields with superior spatial distribution [12]. Consequently, it achieves efficient energy recycling in the time domain and expands the effective interaction volume for nonlinear processes in the spatial domain, ultimately generating a nonlinear synergistic enhancement that surpasses the simple sum of individual modes [13]. Furthermore, QGM resonance exhibit strong robustness against variations in the size of nanoparticles atop the waveguide, posing significant challenges for tuning the resonance $Q$-factor [14,15]. It is well-established that in periodic structures with $C_2$ symmetry, optical responses generated by $x$-polarized and $y$-polarized excitations differ substantially [16]. Currently, tuning the $Q$-factor primarily relies on adjusting the geometric and material parameters of metasurfaces, which inevitably increases the complexity of control [17-19]. Certain modes can be excited by $y$-polarized light but remain inaccessible to $x$-polarization. This implies that the $Q$-factor of hybridized modes can be deliberately tuned by varying the polarization angle of incident light [20].

Nonlinear frequency conversion finds applications across diverse domains including imaging [21], laser technology [22], and quantum photonics [23]. Conventional implementations rely on bulk crystals for nonlinear signal generation, necessitating strict phase-matching conditions to achieve viable conversion efficiency [24]. However, the incorporation of metasurfaces enables relaxation of phase-matching constraints while simultaneously enhancing harmonic conversion efficiency [25,26]. The significant enhancement of THG efficiency fundamentally stems from the exponential intensification of near-field localized light [27-29]. QGM establishes the physical foundation for efficient nonlinear conversion through subwavelength field confinement and broad angular tolerance [30,31]. Mode coupling, meanwhile, overcomes the limitations of single-mode systems via spatiotemporal synergy mechanisms—constructing hybridized electric fields in the spatial domain to expand the effective interaction volume, and leveraging Rabi oscillations in the temporal domain to extend photon lifetime [33,34]. The synergistic interplay between these mechanisms achieves enhanced conversion efficiency and upgraded system robustness, providing a key technological pathway for on-chip tunable nonlinear photonic devices.

This paper investigates the coupling between TE and TM modes excited in a bilayer waveguide and the control of the resonant $Q$-factors of QGMs. By introducing a periodic perturbation of small square cubes above the bilayer waveguide, TE and TM modes are excited into QGMs. Strong coupling between the two modes is achieved by varying the period along the y-direction. For the system entering the strong coupling regime, not only can the $Q$-factor and coupling strength be controlled by adjusting the polarization angle of the incident light without changing material or structural parameters, but the system stability is also enhanced. Furthermore, this strong coupling mechanism can significantly enhance the effect of THG.

## 2. Design and simulations

The proposed metasurface schematics are illustrated in Fig. 1(a), featuring a bilayer waveguide integrated with a periodic square nanocube array. The designed structure is analyzed using the finite element method (FEM). The periodicity along the $x$-direction is fixed at 500 nm. Both silicon (Si) and silicon dioxide (SiO$_2$) slabs maintain a thickness of 400 nm, with refractive indices of 3.47 and 1.46, respectively.

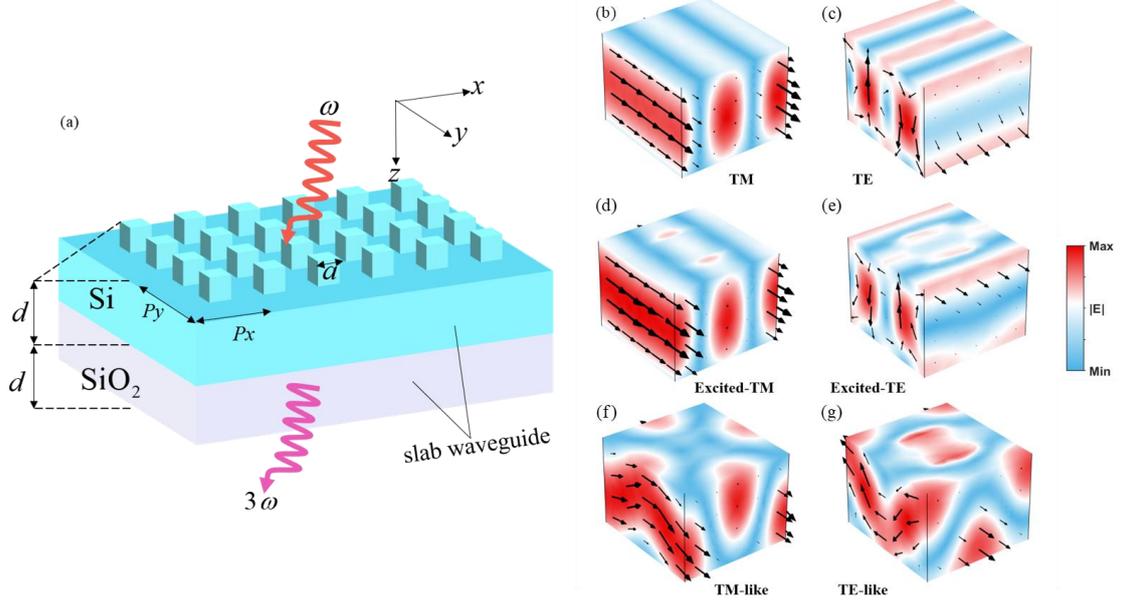

FIG.1 (a) Schematic diagram of the metasurface. The electric field distribution (color) and the electric field vectors (black arrows) of the metasurface (b)-(c) without the cube, and (d)-(g) with the cube $a$ = 223 nm. Both Si and SiO$_2$ slabs have a thickness $d$ of 400 nm. $P_x$ is fixed at 500 nm. In (b)-(e), the $P_y$ is 500 nm, while in (f)-(g), the $P_y$ is 537 nm.

The metasurface supports both TE and TM waveguide modes, where the TM mode is parallel to the incident light polarization and the TE mode is perpendicular to it [34]. Under normal incidence, only light at specific wavelengths satisfying the phase-matching condition can couple into lateral waveguide modes and establish guided resonances. When normally incident light exhibits $y$-direction polarization, the TM and TE mode phase-matching conditions satisfy [35]:

$$\beta_{\mathrm{TM}} = m\frac{2\pi}{P_x}, m = 0, \pm 1, \pm 2, ..., \quad (1)$$

$$\beta_{\mathrm{TE}} = n\frac{2\pi}{P_y}, n = 0, \pm 1, \pm 2, ..., \quad (2)$$

where $\beta_{\mathrm{TM}}$ and $\beta_{\mathrm{TE}}$ are the propagation constants of TM mode and TE mode, respectively, $m$ is the diffraction order of TM wave and n is the diffraction order of TE wave.

## 3. Results and discussion

In order to better explain the physical mechanism, the metasurface without the cube, that is, the Si and SiO$_2$ slabs waveguide is first considered. As shown in Figs. 1(b)-(c), under normal incidence with $y$-direction polarization, TE and TM modes establish standing waves along the $y$- and $x$-directions, respectively. With the introduction of a cubic structure of side length $a$ = 223 nm, both TE and TM modes are excited, exhibiting significantly enhanced field localization around the nanocube (Figs. 1(d)-(e)). At the reference point of $P_y = 537$ nm, the electric fields redistribute and become bent due to mode coupling, as evidenced in Figs. 1(f)-(g). As $P_y$ varies from 500 nm to 555 nm, the TE mode undergoes a redshift, with the TE and TM modes exhibiting minimum frequency separation at $P_y = 537$ nm (Fig. 2(a)). Through multipole decomposition analysis, it is found that at $P_y = 520$ nm, the resonance generated by the TM mode is primarily attributed to the magnetic dipole (MD), electric quadrupole (EQ), and toroidal dipole (TD) terms, while the resonance of the TE mode is mainly dominated by the magnetic dipole (MD) and electric quadrupole (EQ) terms. When $P_y$ is increased to 537 nm, the primary change lies in the radiative power of the multipole term (Figs. 2(b)-(c)).

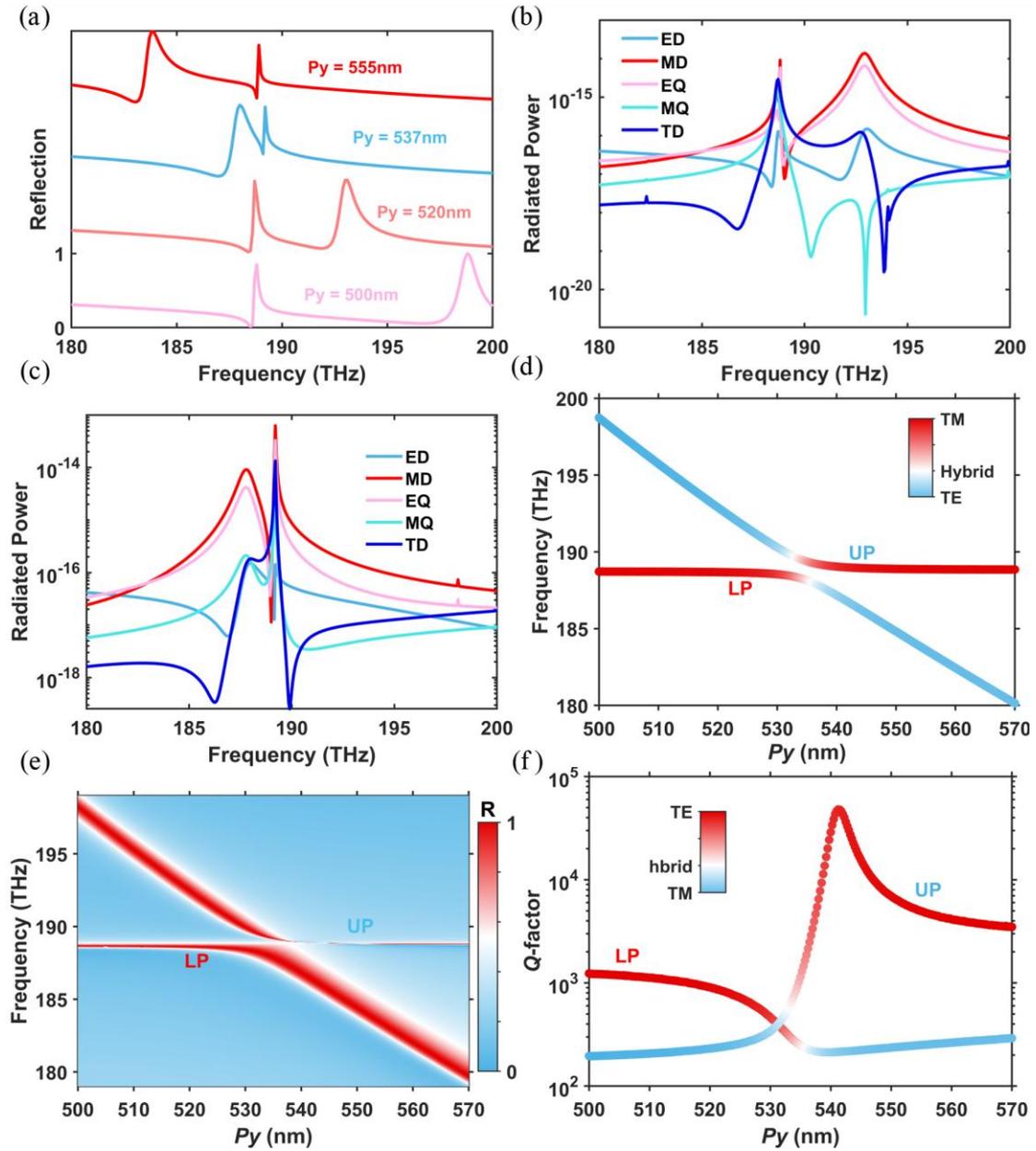

FIG.2 Incident light polarization along the $y$-axis. (a) Reflectance spectra of $P_y$ at 500 nm, 520 nm, 537 nm, and 555 nm. (b)-(c) Multipole decomposition diagrams for $P_y$ at (c) 520 nm and (d) 537 nm. (d) Characteristic frequencies and average polarizability as functions of $P_y$ parameters. (e)-(f) $Q$-factor and reflection spectra dependencies on $P_y$ parameters.

To better explain the dependence of the optical response of the metasurface on the polarization direction, the average polarization state is defined as $(I_{TM}-I_{TE})/(I_{TM}+I_{TE})$, where $I_{TM} = |E_x|^2 + |E_y|^2$ and $I_{TE} = |E_z|^2$ are the intensities of the in-plane (TM) and out-of-plane (TE) components of the electric field, respectively, averaged over one unit cell of the structure [36]. When the resonant frequency of the TE mode approaches that of the TM mode, both the eigenfrequency curves and reflection spectra exhibit anti-crossing behavior, with the mean polarizability approaching zero near the anti-crossing point. This unambiguously demonstrates coupling between the two modes (Figs. 3(d)-(e)). The Hamiltonian for the TE-TM mode coupling mechanism is given by [12]:

$$H = \begin{pmatrix} \omega_{TM} - i\frac{\Gamma_{TM}}{2} & g \\ g & \omega_{TE} - i\frac{\Gamma_{TE}}{2} \end{pmatrix} \quad (3)$$

Here, $\omega_{TM}$ and $\omega_{TE}$ are the resonance eigenfrequencies of uncoupled TM/TE modes, $\Gamma_{TM}$ and $\Gamma_{TE}$ their linewidths, g the intermodal coupling rate. Under the resonance condition ($\omega_{TM} = \omega_{TE} = \omega_0$), the eigenfrequency values of the coupled states can be solved. The eigenfrequency values are:

$$\omega_{UP,LP} = \omega_0 - i\frac{\Gamma_{TM}+\Gamma_{TM}}{4} \pm \sqrt{g^2 - (\frac{\Gamma_{TM}+\Gamma_{TM}}{4})^2} \quad (4)$$

Based on these equations, it is known that there is radiation exchange between these two modes, but the total radiation is conserved. Near the reference point of $P_y = 541$ nm, the reflection bandwidth of the upper branch narrows while that of the lower branch widens. Concomitantly, the $Q$-factor of the upper branch increases, whereas the maximum $Q$-factor of the lower branch decreases, with the upper branch achieving peak $Q$-factors up to $6\times10^4$. This indicates that energy exchange occurs between the two modes, resulting in enhanced dissipation in the lower polariton branch and reduced dissipation in the upper polariton branch. The Rabi splitting $\Omega_R$ between two anti-crossing energy levels can beobtained as,

$$\hbar\Omega_R = 2\hbar\sqrt{g^2 - (\Gamma_{TM}-\Gamma_{TE})^2/16} \quad (5)$$

In the strong coupling regime, the Rabi splitting is larger than the mean linewidth of the two coupled resonant modes,

$$\Omega_R > \frac{\Gamma_{TM}+\Gamma_{TM}}{2} \quad (6)$$

The calculated Rabi splitting energies are $\hbar\Omega_R = 2.6$ meV, $\hbar\Omega_{TM} = 0.37$ meV, and $\hbar\Omega_{TE} = 2.15$ meV. Since the Rabi splitting energy $\hbar\Omega_R$ exceeds the average of $\hbar\Omega_{TM}$ and $\hbar\Omega_{TE}$, this indicates that the system is in the strong coupling regime.

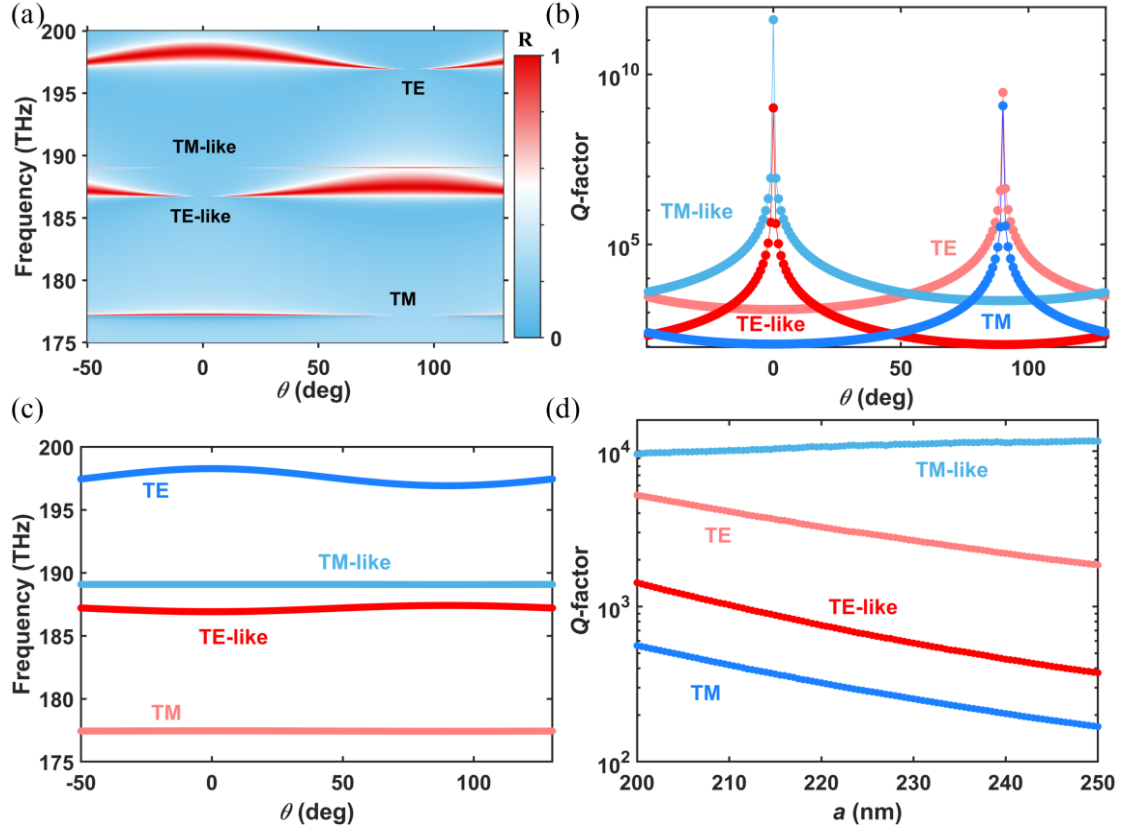

FIG.3 (a)-(c) Evolution of reflection spectra, $Q$-factors and characteristic frequencies for TE, TM, TE-like, and TM-like modes, respectively, as a function of incident light polarization angle. (d) $Q$-factor dependence on small cube side length. All subfigures incorporate a small dielectric cube. Parameters: Side length $a = 223$ nm in (a)-(c); Polarization angle $\theta = 35°$ in (d).

When the incident light polarization is aligned along the x-axis, $\theta = 0$. $P_y$ is set to 537 nm. After the system enters the strong coupling regime, the structure exhibits $C_2$ symmetry. This results in distinct optical response characteristics for polarizations along the *x*-axis and *y*-axis. Figure 3(a)-(b) reveal that as the polarization angle ($\theta$) of incident light varies, the reflection spectra of the TE and TM modes exhibit their narrowest linewidths (corresponding to maximal $Q$-factors) at $\theta = 90°$, while the TE-like and TM-like modes reach their minimal linewidths (peak $Q$-factors) at $\theta = 0°$. This indicates that tuning the incident light polarization angle simultaneously modulates both the $Q$-factors. At $\theta = 0°$, the TM-like mode achieves a $Q$-factor of up to $10^{12}$ – two orders of magnitude higher than the maximum $Q$-factors of the other three modes – while maintaining polarization-independent characteristic frequencies. The resonance frequency and $Q$-factor of the TE-like mode are tunable via the incident light's polarization angle, thus enabling control of the coupling strength. Furthermore, the $Q$-factor of the TM-like mode exhibits greater stability against variations in the side length of the small cube compared to the other three modes (Figs. 3(b)-(d)). Therefore, systems incorporating the TM-like mode enable not only dynamic control of the $Q$-factor and coupling strength, but also enhanced structural stability.

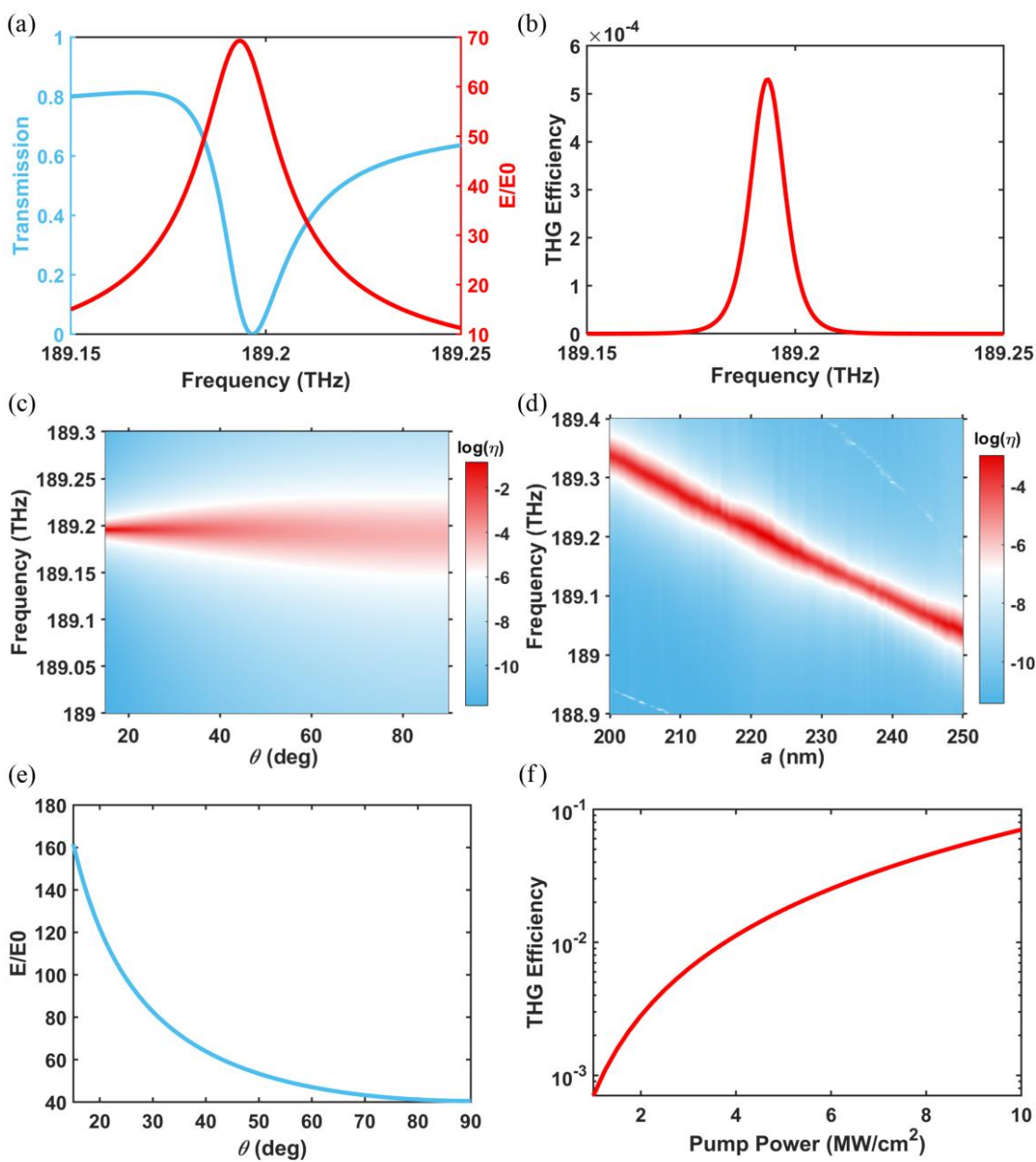

FIG.4 (a) is the plot of transmission and maximum field enhancement for the fundamental wave. (b)-(d) and (f) are the efficiency plots for the third harmonic. (e) is the plot showing the relationship between the incident light polarization angle and the maximum field enhancement. (a)-(f) all involve the introduction of a small cube. In (a), (b), (d) and (f), $\theta$ = 35°. In (a)-(c) and (e)-(f), $a$ = 223 nm.

To verify the performance of this coupled system in nonlinear frequency conversion, the incident light polarization angle was controlled to excite a high-$Q$ QGM, with the aim of significantly enhancing the third-harmonic generation efficiency. The third harmonic is generated by the nonlinear

polarization effect of silicon, which can be represented by the following formula for nonlinear polarization:

$$\vec{P}(3\omega) = 3\varepsilon_0 \chi^3 (\vec{E}(\omega) \cdot \vec{E}(\omega)) \vec{E}(\omega) \tag{7}$$

Here, $\varepsilon_0$ refers to the permittivity of free space, while $\chi^{(3)} = 2.45 \times 10^{-19}$ V/m² represents the third-order nonlinear optical coefficient of silicon material. The efficiency of third-harmonic generation can be expressed using the formula $\eta = P_{THG} / P_{FF}$, where $P_{THG}$ is the power of the third harmonic and $P_{FF}$ is the power of the pump source [16].

In the analysis of third-harmonic generation efficiency, $P_y$ is fixed at 537 nm. Figure 4 (a) indicates that the frequency of the pump source is 189.19 THz, and the maximum field enhancement can reach 70 times that of the incident light. At this frequency, the efficiency of third-harmonic generation can reach $5 \times 10^{-4}$ (Fig. 4(b)). As the polarization angle of the incident light approaches 15 degrees, the efficiency of third-harmonic generation can exceed $10^{-2}$, and the maximum field enhancement can also reach 160 times that of the incident light (Figs. 4 (c) and (e)). Owing to the robustness of this coupled sys observed that the third-harmonic generation efficiency remains stable at approximately $5 \times 10^{-4}$, even as parameter 'a' varies between 200 and 250 (Fig. 4(d)). Figure 4(f) demonstrates the relationship between the efficiency of third-harmonic generation and the pump power. When the pump power deviates slightly from the initial pump power, the efficiency shows a predominantly linear increase with respect to the pump power. From this result, we can infer that this metasurface not only has a high third-harmonic efficiency in experiments but can also suppress the impact of manufacturing defects on efficiency.

## 4. Conclusion

In summary, through energy exchange between two modes in a strongly coupled system, the loss of one mode is reduced, thereby significantly enhancing its $Q$-factor. Furthermore, by varying the polarization angle of the incident light, the system's $Q$-factor and coupling strength can be modulated. Within this system, the maximum $Q$-factor can reach $10^{12}$. In previous extensive research on $Q$-factor modulation of resonances in metasurfaces, structural or material parameter alterations were typically required, which significantly increased the difficulty of control. In contrast, our approach only requires adjusting the polarization of the incident light to achieve effective control of the $Q$-factor, without changing structural or material parameters. When applying this structure to third-harmonic generation, a conversion efficiency of $10^{-2}$ can be achieved.

**Funding.** Hunan Provincial Natural Science Foundation of China (2024JJ5366, 2021JJ40523, 2024JJ5094), Scientific Research Foundation of Hunan Provincial Education Department (22B0105), and National Natural Science Foundation of China (62205278, 11947062).